\documentclass[12pt]{iopart}

\usepackage[english]{babel}
\usepackage[utf8x]{inputenc}
\usepackage[T1]{fontenc}

\usepackage{iopams} % import useful math packages
\usepackage{xcolor}
\usepackage[colorlinks,allcolors=black]{hyperref}
\usepackage{graphicx}
\usepackage{booktabs}
\usepackage[acronym]{glossaries-prefix}
\usepackage{placeins}
\usepackage[print-unity-mantissa=false]{siunitx}
\usepackage{mathtools}

\DeclareMathOperator{\sinc}{sinc}
\DeclareMathOperator{\rect}{rect}
\DeclareMathOperator{\sgn}{sgn}
\DeclareMathOperator*{\argmin}{arg\,min}
\DeclareMathOperator{\E}{E}

\newcommand{\delay}[1]{\mathbf{D}_{#1}}
\newcommand{\fdelay}[1]{\mathcal{D}_{#1}}
\newacronym[prefixfirst={the\ }]{lisa}{LISA}{Laser Interferometer Space Antenna}
\newacronym{gw}{GW}{gravitational wave}
\newacronym{esa}{ESA}{European Space Agency}
\newacronym{tdi}{TDI}{time-delay interferometry}
\newacronym{fir}{FIR}{finite impulse response}
\newacronym{lti}{LTI}{linear time-invariant}
\newacronym[longplural=power spectral densities]{psd}{PSD}{power spectral density}
\newacronym[longplural=amplitude spectral densities]{asd}{ASD}{amplitude spectral density}
\newacronym[longplural=moveable optical sub-assemblies]{mosa}{MOSA}{moveable optical sub-assembly}
\newacronym{prn}{PRN}{pseudorandom noise}

\begin{document}

\title[Optimal design of interpolation methods for TDI]{Optimal design of interpolation methods for time-delay interferometry}

\author{Martin Staab$^{1,2}$, Jean-Baptiste Bayle$^3$, Olaf Hartwig$^4$, Aur\'elien Hees$^1$, Marc Lilley$^1$, Graham Woan$^3$, and Peter Wolf$^1$}

\address{$^1$ SYRTE, Observatoire de Paris, Universit\'e PSL, CNRS, Sorbonne Universit\'e, LNE, Paris, France}
\address{$^2$ Institute for Gravitational and Subatomic Physics (GRASP), Department of Physics, Utrecht University, Princetonplein 1, NL-3584 CC Utrecht, The Netherlands}
\address{$^3$ University of Glasgow, Glasgow G12 8QQ, United Kingdom}
\address{$^4$ Max Planck Institute for Gravitational Physics (Albert Einstein Institute), Hannover, Germany}

\ead{m.b.staab@uu.nl}
% \vspace{10pt}
% \begin{indented}
% \item[]November 2024
% \end{indented}

\begin{abstract}
\Gls{tdi} suppresses laser frequency noise by forming linear combinations of time-shifted interferometric measurements. The time-shift operation is implemented by interpolating discretely sampled data. To enable in-band laser noise reduction by eight to nine orders of magnitude, interpolation has to be performed with high accuracy. Interpolation can be understood as the convolution of an interpolation kernel with the data to be shifted. Optimizing the design of this interpolation kernel is the focus of this work. Previous research that studied constant time-shifts suggested Lagrange interpolation as the interpolation method for \gls{tdi}. Its transfer function is maximally flat at DC and therefore performs well at low frequency. However, to be accurate at high frequencies, Lagrange interpolation requires a high number of coefficients. Furthermore, when applied in \gls{tdi} we observed prominent time-domain features when a time-varying shift scanned over a pure integer sample shift. To limit this effect we identify an additional requirement for the interpolation kernel: when considering time-varying shifts the interpolation kernel must be sufficiently smooth to avoid unwanted time-domain transitions that produce glitch-like features in power spectral density estimates. The Lagrange interpolation kernel exhibits a discontinuous first derivative by construction, which is insufficient for the application to \acrshort{lisa} or other space-based \gls{gw} observatories. As a solution we propose a novel design method for interpolation kernels that respect a predefined requirement on in-band interpolation residuals and that possess continuous derivatives up to a prescribed order. Using this method we show that an interpolation kernel with 22 coefficients is sufficient to respect \acrshort{lisa}'s picometre-requirement and to allow for a continuous first derivative which suppresses the magnitude of the time-domain transition adequately. The reduction from 42 (Lagrange interpolation) to 22 coefficients enables us to save computational cost and increases robustness against artefacts in the data.
\end{abstract}

\submitto{\CQG}

\clearpage

\glsresetall
\section{Introduction}
% general introduction to LISA

%TODO: cite reference for ground-based detectors and PTAs
\Pgls{lisa} is an ESA-led space mission with a planned launch date in the mid-2030s to detect \glspl{gw} in the frequency band from \qty{0.1}{\milli\Hz} to \qty{1}{\Hz}~\cite{Colpi:2024xhw}, thereby filling the spectral gap between pulsar timing arrays and ground-based detectors like LIGO, Virgo and Kagra. The millihertz-regime is densely populated with a myriad of \gls{gw} sources. The detection of those will shed new light on black hole formation history, test the strong-field gravity regime and trace \glspl{gw} across cosmic ages~\cite{LISA:2017pwj} (for more information on the astrophysics with \pgls{lisa} see~\cite{LISA:2022yao} and references therein). 

\Pgls{lisa} consists of three spacecraft separated by approximately \num{2.5} million kilometres forming a near-equilateral triangle. Each satellite carries two \glspl{mosa} that each host a laser source, an optical bench and a free-falling test-mass. Laser interferometry is employed to measure picometre-scale changes between those test-masses that serve as inertial reference points. The sensitivity of \pgls{lisa} is limited by spurious accelerations of the test masses at low frequencies, and by shot noise in the readout of the interferometric beatnotes at high frequency~\cite{Colpi:2024xhw}.

Instabilities of the central terahertz-frequency of the lasers necessitate ranging along at least two baselines. Comparison of the two ranging observables enables suppression of laser frequency noise. Ground-based detectors achieve this suppression by constructing Michelson interferometers with arms of equal length, so that the laser noise is cancelled upon recombination of the two beams. In space, test masses that represent the end-mirrors of the interferometers follow their individual orbital trajectories. Therefore, the triangle formed by the three \pgls{lisa} spacecraft is non-rigid, and is constantly flexing, so the light travel times along the six links differ by a few percent and fluctuate over the course of year with a typical rate of \qty{3e-8}{\s\per\s}~\cite{Martens:2021phh}. This renders the simple Michelson interferometer ineffective.

Furthermore, physically realizing round-trips to form equal-arm interferometers in space is impractical. Retro-reflection at the end-stations is unfeasible as laser beams diverge, and only a small fraction of the power (a few hundred picowatts~\cite{Colpi:2024xhw}) is received by the distant satellite.  The reflected beam would be reduced by the same factor on the return leg, down to immeasurable levels.
%Furthermore, the construction of higher-generation topologies in space, involving multiple round-trips, would greatly (and unnecessarily) increase the complexity of the instrument.

The solution is to rely on single-link measurements and construct virtual equal-arm interferometers in post-processing, on the ground. The \gls{tdi} algorithm achieves this by time-shifting and linearly combining the single-link measurements. \Gls{tdi} enables realization of numerous interferometer topologies, e.g., Michelson-type or Sagnac-type~\cite{Vallisneri:2005ji,Tinto:2020fcc}. Due to \pgls{lisa}'s unequal and flexing arms ''higher-generations`` of those topologies need to be considered, where the two virtual beams take multiple round-trips across the constellation to reduce the difference in travelled path length. In practice, the so-called second-generation variables, that suppress laser frequency noise up to first order in inter-spacecraft velocities and zeroth order in accelerations, are adequate for the \pgls{lisa} mission~\cite{Shaddock:2003dj,Cornish:2003tz}.

The performance of \gls{tdi} is limited by several technical laser noise residuals which enter at various stages in data processing~\cite{Staab:2023qrb}. On board the spacecraft, filters and decimators give rise to the flexing-filtering effect~\cite{Bayle:2018hnm} and produce aliased laser noise. As shown in~\cite{Staab_phd_thesis} the flexing-filtering effect can be mitigated, however any aliased laser noise remains in the data as an uncorrelated noise contribution. When performing \gls{tdi} on the ground, ranging noise and interpolation errors couple to the final \gls{tdi} variables. To reduce the effect of ranging noise, dedicated processing pipelines are being designed that combine information from different ranging sensors~\cite{Reinhardt:2023ccg}. However, the impact of interpolation errors has not received much attention, and is therefore the subject of this paper.

\Gls{tdi} requires high-precision interpolation to effectively suppress laser frequency noise by up to nine orders of magnitude. Phase measurements taken on-board are sampled at a few Hertz and sent to the ground. To time-shift the data and construct \gls{tdi} combinations in post-processing the time series are interpolated and re-evaluated on the shifted time grid~\cite{Shaddock:2004ua}. The relative precision of the interpolation method must be of the order of \num{1e-9} to respect the laser noise suppression ratio. Lagrange interpolation was identified to outperform other methods (windowed sinc) using the fewest number of interpolation coefficients~\cite{Shaddock:2004ua}. Therefore, efficient algorithms to calculate those were developed and traditionally used in data processing pipelines for space-based \gls{gw} observatories~\cite{halloin2018}.

However, in this paper, we will show that Lagrange interpolation is suboptimal for two reasons. First, when considering realistic \pgls{lisa} data involving time-dependent time-shifts, the interpolation error becomes non-stationary. This produces prominent time-domain features which are connected to sudden changes in the interpolation coefficients. For Lagrange interpolation, this happens when the time-shift passes through an integer number of sampling periods as the interpolation coefficients adjust in a non-smooth fashion. This property was missed in previous studies, as only constant time-shifts were considered. Second, we find that Lagrange interpolation uses a relatively large number of coefficients which is requires more computational resources and increases the amount of invalid samples at the boundaries of the time series.

To remedy both shortcomings, we will present a novel interpolation method. It allows for specifying the smoothness of the kernel function, i.e. the function from which the interpolation coefficients are drawn. We find that a continuous first derivative suffices to suppress the time-domain feature to adequate levels. Furthermore, we describe an optimization routine to minimize the number of interpolation coefficients while retaining sufficient laser noise suppression over the entire \pgls{lisa} band.

The paper is organized as follows. In \sref{sec:tdi} we introduce the basic concept of \gls{tdi} and explain the coupling of interpolation errors. Then, in \sref{sec:interpolation} we develop a general framework to describe interpolation as a convolution of a kernel function and the data. We derive the kernel function for Lagrange interpolation in \sref{sec:lagrange_interpolation} and present its shortcomings. As a solution, in \sref{sec:cosine-sum_kernel} we introduce a novel class of kernel functions dubbed cosine-sum kernels. Finally, in \sref{sec:simulation} we validate our findings for both interpolation methods by running numerical simulations, and conclude in \sref{sec:conclusion}.

\section{Interpolation errors in TDI}
\label{sec:tdi}
% introduce the LISA case and why the interpolation error matters
% motivation of the problem

\begin{figure}
    \centering
    \includegraphics{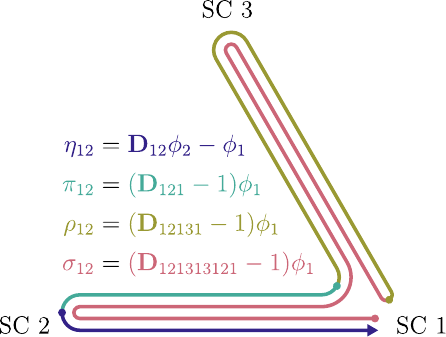}
    \caption{Illustration of the working principle of \gls{tdi}. Increasingly longer virtual photon paths are constructed by hierarchically building up the intermediary variables $\pi$, $\rho$ and $\sigma$ as defined in equations \ref{eq:pi}, \ref{eq:rho} and \ref{eq:sigma}. In the final step the two round-trip variables $\sigma_{12}$ and $\sigma_{13}$ are subtracted to form the second-generation Michelson variable $X_2$. The two synthesized counter-propagating photon path have almost identical length, hence, laser noise is greatly suppressed. Adopted from \cite{Staab:2023qrb}.}
    \label{fig:intervars}
\end{figure}

% describe LISA MOSAs: laser, test-mass, interferometers
\pgls{lisa} does not physically realize equal-arm interferometers in space. Instead, it relies on the so-called split-interferometry setup which employs three interferometers on each of the six optical benches: the inter-spacecraft, the reference and the test-mass interferometer. The inter-spacecraft interferometer mixes light from the distant laser that has travelled to the local optical bench, thereby integrating the \gls{gw} strain along the line-of-sight. In the reference interferometer, the adjacent laser is compared to the local laser via the backlink fibre. Finally, the test-mass interferometer is almost identical to the reference interferometer with the difference that the adjacent light is bounced off the test mass, which serves as an inertial reference for the ranging measurement.~\cite{Colpi:2024xhw}

% make comment that we only keep laser phases/no other noises as laser noise is by far the domininant noise when interpolating measurements and drives the requirements for the interpolation methodess to WLCG or collaboration resour
% also construction of intermediary already produces interpolation errors
The aforementioned interferometers are then combined to form the six single-link measurements $\eta_{ij}$~\cite{Tinto:2020fcc}. First, the inter-spacecraft and test-mass interferometers are combined to compensate for the longitudinal relative motion between the optical bench and the test-mass. Then, the reference interferometers are brought in to reduce the number of lasers to one per spacecraft. The single-link measurements, in terms of laser phases $\phi_i$, are therefore given by
\begin{equation}
    \eta_{ij}(t) = \delay{ij} \phi_j(t) - \phi_i(t) . \label{eq:eta}
\end{equation}
Here, $i$ denotes the local and $j$ the distant spacecraft index. The time-delay operator $\delay{ij}$ models the propagation of the distant laser phase $\phi_j$ to the local spacecraft by delaying it by the pseudo-range $d_{ij}(t)$ expressed as time. We formally define the general delay operation as
\begin{equation}
    \delay{} x(t) = x(t - d(t)) . \label{eq:delay_operator}
\end{equation}
In \pgls{lisa}, the time-delay $d(t)$ contains the ''macroscopic`` light travel time of about eight seconds, the result of desynchronization between time reference frames (relativistic effects and clock imperfections) and the path-length variations due to \glspl{gw}. Dedicated ranging sensors are implemented in \pgls{lisa} to measure all six delays $d_{ij}$ independently of the carrier-carrier beatnote.

The ultimate goal of \gls{tdi} is to suppress laser noise by constructing virtual equal-arm interferometers; in our example, a simple Michelson interferometer. In the first step we cancel the distant laser from $\eta_{ij}$ by constructing round-trip interferometers as illustrated in \fref{fig:intervars}. This is achieved by adding a delayed version of the distant readout $\eta_{ji}$ as
\begin{align}
    \pi_{ij} &= \eta_{ij} + \fdelay{ij}\eta_{ji} \\
    &= (\delay{iji} - 1) \phi_i + \underbrace{(\fdelay{ij} - \delay{ij})}_{\bDelta_{ij}} \eta_{ji} . \label{eq:pi}
\end{align}
Here, we introduce the so-called ''post-processing`` delay $\fdelay{}$ that is applied on the ground. As the raw interferometric measurements are discretely sampled at a few hertz, we rely on interpolation to apply arbitrary time-shifts in \gls{tdi}. Therefore, the post-processing delay operator $\fdelay{}$ represents only an approximation of the true delay $\delay{}$ defined in \eref{eq:delay_operator}. Any deviation of the post-processing from the true delay operator is described by the interpolation error $\bDelta_{ij}$. In reality, the post-processing delay operator is also impacted by other imperfection, e.g., ranging biases~\cite{Staab:2023qrb}. However, here we only focus on the residual laser noise in the final \gls{tdi} variables resulting from interpolation errors.

In \eref{eq:pi} we introduce the short-hand notation for nested delays. In general, we define the contracted delay operator
\begin{equation}
    \delay{i_1 \cdots i_N} \equiv \delay{i_1 i_2} \cdots \delay{i_{N-1} i_N} , \label{eq:contracted_delay}
\end{equation}
where the associated nested time-delay $d_{i_1 \cdots i_N}(t)$ can be calculated recursively. The equivalence in \eref{eq:contracted_delay}, however, does not hold for the post-processing delay operator $\fdelay{}$ as each application introduces additional interpolation errors. Consequently, the results of $\fdelay{i_1 \cdots i_N} x(t)$ and $\fdelay{i_1 i_2} \cdots \fdelay{i_{N-1} i_N} x(t)$ generally differ.

Now, we take the difference of two adjacent round-trip measurements and yield the simple Michelson interferometer,
\begin{equation}
    X_0 = \pi_{13} - \pi_{12}  = \left(\delay{131} - \delay{121}\right) \phi_1 + (\bDelta_{13} \eta_{31} + \bDelta_{12} \eta_{21}) . \label{eq:x0}
\end{equation}
In the equal-arm assumption, laser noise from the first term cancels, leaving only laser noise residuals due to interpolation errors from the second term. In a realistic \pgls{lisa} setup inter-spacecraft delays are unequal and time-varying. Therefore, the second-generation Michelson variable $X_2$ is required to achieve nearly equal arms and suppress laser noise effectively (see \fref{fig:intervars} and also~\cite{Vallisneri:2005ji}). Let us define the (intermediary) variables as in \cite{Staab:2023qrb}
\begin{align}
    \rho_{ij} &= \pi_{ij} + \fdelay{iji}\pi_{ik} , \label{eq:rho}\\
    \sigma_{ij} &= \rho_{ij} + \fdelay{ijiki}\rho_{ik}, \label{eq:sigma}\\
    X_2 &= \sigma_{13} - \sigma_{12} \approx [[\delay{131}, \delay{121}], \delay{12131}] \phi_1 + \delta\!X_2^{\fdelay{}} . \label{eq:x2}
\end{align}
Here, the delay operator difference in the last line is expressed as nested commutators to emphasize that the fundamental laser noise residual is limited by contributions quadratic in velocities $\dot d_{iji}$ and linear in accelerations $\ddot d_{iji}$. The interpolation residual $\delta\!X_2^{\fdelay{}}$ collects terms from the various intermediate variables\footnote{The index $k$ appearing in equations \ref{eq:rho}, \ref{eq:sigma} and \ref{eq:x2} denotes the remaining spacecraft index that is different from $i$ and $j$.} $\pi_{ij}$, $\rho_{ij}$ and $\sigma_{ij}$~\cite{Staab:2023qrb}. Here, we make the particular choice of applying the nested post-processing delays $\fdelay{iji}$ ($\neq \fdelay{ij} \fdelay{ji}$) and $\fdelay{ijiki}$ ($\neq \fdelay{ij}\fdelay{ji}\fdelay{ik}\fdelay{ki}$) for the sake of computational efficiency. The implications of this factorization are further discussed in~\cite{Staab:2023qrb}. Moreover, we neglect second-order couplings in $\delta\!X_2^{\fdelay{}}$, i.e. couplings of interpolation errors to interpolation residuals, as the latter is already small compared to the laser noise content.

\section{The interpolation kernel}
\label{sec:interpolation}

\begin{figure}
    \centering
    \includegraphics{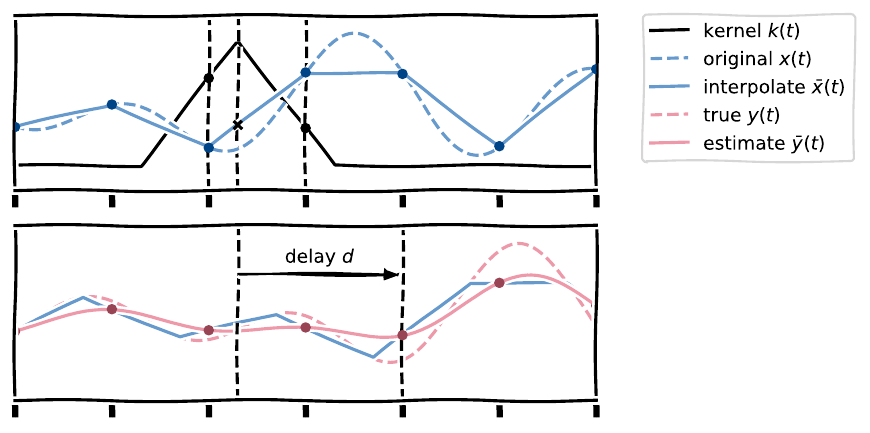}
    \caption{Illustration of the post-processing delay operation. In the upper panel, the solid blue line depicts a simple linear interpolation between discrete samples (blue dots) of the continuous band-limited signal $x(t)$ (dashed blue line). The black line shows the corresponding triangular interpolation kernel. To interpolate at a given time (black cross) the neighbouring samples are weighted by the corresponding readings of the kernel function (black dots). In the lower panel we plot the time-delayed interpolated signal in blue, its sampled version (the result of \eref{eq:delayed}) as red dots and its continuous-time band-limited equivalent (red line). In dashed red we overlay the true delayed signal. We observe an amplitude and phase error between the post-processing delay operation and the true delay operation.}
    \label{fig:scheme}
\end{figure}

% explain concept of local interpolation: use set of point around time of interest -> computational advantage: locality and inexpensive
% interpolating equally-spaced samples in time
% weighting samples by interpolation kernel $k(t)$ which depends on the distance between time $t$ and time of point
% $x(t)$ is a band-limited signal (only frequency content up to Nyquist frequency)
Let us now turn to the actual implementation of the post-processing delay $\fdelay{}$ for \gls{tdi}. The operation is performed in two steps, summarized in \fref{fig:scheme}. First, the discrete-time samples $x(n T_\mathrm{s})$ (illustrated as blue dots) of the original time series are interpolated. Here, $T_\mathrm{s}$ denotes the sampling interval between samples. This is achieved by taking the weighted sum of samples close to the time of interest achieving locality of the operation. The weights are drawn from the so-called interpolation kernel which is shown in black. Formally we define the continuous-time interpolated signal as
\begin{equation}
    \bar x(t) = \sum_{m=-\infty}^\infty x(m T_\mathrm{s}) \cdot k(t - m T_\mathrm{s}) . \label{eq:interpolation}
\end{equation}
The choice of kernel function determines the quality of the interpolation. For perfect band-limited interpolation one would need to set $k(\tau) = \sinc(f_\mathrm{s} \tau) = \frac{\sin(\pi f_\mathrm{s} \tau)}{\pi f_\mathrm{s} \tau}$ which yields the Whittaker-Shannon interpolation formula~\cite{jenkins_spectral_1968}. The critical downside of this approach is that the $\sinc$-kernel is infinitely wide, so the sum in \eref{eq:interpolation} runs over an infinite number of elements. To make the computation feasible, we only consider kernels of finite width $T$. Furthermore, we limit ourselves to the class of real and even functions, i.e. symmetric in $\tau$ and possessing a real and even Fourier transform, which becomes important later.

%  - for practial reasons width (where kernel is non-zero) is equal to $N T_s$ where $N$ is an integer

In the next step, we apply the delay $d$ to the interpolated time series $\bar x(t)$ and sample it on the original time grid. Formally, This translates to
\begin{align}
    \bar y\!\left(n T_\mathrm{s}\right) &= \bar x(t - d) \big|_{t=n T_\mathrm{s}} ,\\
    &= \sum_{m=-\infty}^\infty x\!\left(m T_\mathrm{s}\right) \cdot k\!\left((n - m) T_\mathrm{s} - d\right) ,\\
    &= \sum_{m=-\infty}^\infty x\!\left((n - m) T_\mathrm{s}\right) \cdot k\!\left(m T_\mathrm{s} - d\right) , \label{eq:delayed}
\end{align}
which yields the delayed time series depicted as red dots in \fref{fig:scheme}. In the last line we performed the substitution $m \rightarrow n - m$ to identify the delay operation as a discrete convolution of the original time series $x(n T_\mathrm{s})$ and the sampled delayed kernel function $k(n T_\mathrm{s} - d)$.

%TODO: better say that we go to continuous time and we sample later -> causes aliasing, but there is none because if x is band-limited than y is also
In the following, we reformulate the post-processing delay operation in continuous time by using the Whittaker-Shannon interpolation formula. This enables us to identify the transfer function of the operation, which can be used to assess its performance. The Whittaker-Shannon formula states that a time series $x(t)$ that is band-limited up to some frequency $f_\mathrm{s} / 2$ can be represented by discrete samples as
\begin{equation}
    x(t) = \sum_{n=-\infty}^\infty x\!\left(n T_\mathrm{s}\right) \cdot \sinc\!\left(f_\mathrm{s} t - n\right) . \label{eq:whittaker-shannon}
\end{equation}
We use this formula to identify the continuous-time representation of $\bar y(m T_\mathrm{s})$ and find that the post-processing delay operation can indeed be represented as a \gls{lti} system for constant time-shifts
\begin{align}
    \bar y(t) &= \sum_{n=-\infty}^\infty \bar y\!\left(n T_\mathrm{s}\right) \cdot \sinc\!\left(f_\mathrm{s} t - n\right) , \\
    &= \sum_{m,n=-\infty}^\infty x\!\left((n - m) T_\mathrm{s}\right) \cdot k\!\left(m T_\mathrm{s} - d\right) \cdot \sinc\!\left(f_\mathrm{s} t - n\right) , \\
    &= \int_{-\infty}^\infty\! \underbrace{k(\tau - d) \sum_{m=-\infty}^\infty \delta\!\left(\tau - m T_\mathrm{s}\right)}_{h(\tau;d)} \cdot x(t - \tau) \,\mathrm{d}\tau ,
\end{align}
where we have introduced the integral over the Dirac-delta function to identify $h(\tau;d)$ as the impulse response of the operation given the delay $d$. Taking the Fourier transform of the latter lets us derive the transfer function $\tilde h(f;d)$ as
\begin{align}
    \tilde h(f;d) &= \sum_{m=-\infty}^\infty k\!\left(m T_\mathrm{s} - d\right) \cdot e^{-2\pi i f m T_\mathrm{s}} , \\
    &= f_s \sum_{m=-\infty}^\infty \tilde k\!\left(f - m f_\mathrm{s}\right) \cdot e^{-2\pi i \left(f - m f_\mathrm{s}\right) d} .
\end{align}
Here, we list two equivalent forms\footnote{The first expression is obtained by inserting $h(\tau;d)$ into the definition of the Fourier transform while the second line is obtained by using the fact that $h(\tau;d)$ is the convolution of $k(t)$ with a Dirac comb whose Fourier transform is also a Dirac comb.}: The first one is convenient for numerical calculations as the sum can be simplified to only run over a finite number of indices when considering kernels of finite width. Furthermore, the kernel function $k(t)$ is readily available to draw the required samples $k\!\left(m T_\mathrm{s} - d\right)$. The second representation can be understood as the spectral folding (similar to aliasing) of the Fourier transform of the kernel multiplied with the transfer function of the delay operation, i.e. the complex exponential $e^{-2 \pi i f d}$. We note that any components of $\tilde k(f)$ residing outside of the band-limit (i.e. for frequencies $|f| > f_\mathrm{s} / 2$) give rise to a phase error in the post-processing delay operation.

\subsection{Interpolation error}
Let us study the amplitude and phase error of the post-processing constant delay operation more systematically. We define the interpolation error as the deviation of the post-processing delay operation from the true delay operation
\begin{equation}
    \bar y(t) - y(t) = \underbrace{\fdelay{} x(t)}_{(h * x)(t)} - \underbrace{\delay{} x(t)}_{x(t - d)} . \label{eq:interpolation_error}
\end{equation}
Let $x(t)$ be a band-limited random process with \gls{psd} $S_x(f)$. Then, we can characterize the interpolation error by calculating its \gls{psd}. As the operation $\bDelta = \fdelay{} - \delay{}$ is a \gls{lti} system itself we can readily write down the result which is given as the product of the magnitude squared of its transfer function and the original \gls{psd} (see e.g.~\cite{Staab:2023qrb})
\begin{align}
    S_{\!\bDelta}(f) &= \big|\underbrace{\tilde h(f;d) - e^{-2 \pi i f d}}_{\tilde{\bDelta}}\big|^2 \cdot S_x(f) , \label{eq:interpolation_error_psd} \\
    &\le \left(\left|f_\mathrm{s} \tilde k(f) - 1\right| + \sum_{m \neq 0} \left|f_\mathrm{s} \tilde k\!\left(f - m f_\mathrm{s}\right)\right|\right)^2 \cdot S_x(f) . \label{eq:worst_case_interpolation_error}
\end{align}
In the second line we have additionally derived an upper bound for the result which is independent of the delay $d$. Here, we have used the triangle inequality to move the norm $|\cdot|^2$ under the sum. From this expression we can derive the condition for the ideal interpolation kernel. For the upper limit to vanish, the Fourier transform of the ideal interpolation kernel must be equal to $\frac{1}{f_\mathrm{s}}$ up to the Nyquist frequency and equal to zero past it; which describes the scaled rectangular function that is the Fourier transform of the $\sinc$-kernel. As discussed previously, the $\sinc$-kernel is not an appropriate choice as it has infinite width. Therefore, in sections \ref{sec:lagrange_interpolation} and \ref{sec:cosine-sum_kernel} we discuss two possible interpolation methods with kernel functions of finite width; Lagrange interpolation and cosine-sum kernels.

\subsection{Interpolation glitch}

Until now, we have only considered constant delays. However, in \pgls{lisa} inter-spacecraft distances are time-dependent with maximum rates of \qty{3e-8}{\s\per\s}~\cite{Martens:2021phh}. Therefore, \gls{tdi} also has to apply time-varying post-processing delay operations to suppress laser frequency noise effectively~\cite{Tinto:2003vj}. As a result the interpolation error derived in \eref{eq:interpolation_error} becomes non-stationary and can produce prominent glitch-like time-domain transitions (see \fref{fig:lisa_glitch} as an example) if the interpolation kernel is not carefully designed.

% explain intuitive understanding
The overall smoothness of the kernel function $k(t)$ determines the amplitude and the time of occurrence of the interpolation glitch. Let us take the example of linear interpolation which is presented in \fref{fig:scheme}. The kernel for linear interpolation is given by the triangular function whose first derivative is discontinuous at $t \in \left\{-T_\mathrm{s}, 0, T_\mathrm{s}\right\}$. Those discontinuities are inherited by the interpolated function $\bar x(t)$ at $t = n T_\mathrm{s}$ where $n \in \mathbb{Z}$. Therefore, when the delay becomes an integer multiple of the sampling time $T_\mathrm{s}$ the delayed signal $\bar y(n T_\mathrm{s})$ scans over such a discontinuity. As a result, the interpolation error changes abruptly and appears to be incoherent before and after the critical time. 

We can estimate the impact of this effect by studying the spectral content of a finite stretch of the delayed time-series $\bar y(t)$ around a critical point $t_0$ which corresponds to a critical delay $d_0 = d(t_0)$ where some derivative $k^{(q)}(m T_\mathrm{s} - d_0)$ in \eref{eq:delayed} is discontinuous (in the case of linear interpolation it is the first derivative). We assume that the delay $d(t)$ changes slowly with time, hence, we can approximate it as a linear function in the vicinity of $t_0$,
\begin{equation}
    d(t) = d_0 + \dot d \cdot \left(t - t_0\right) .
\end{equation}
Then, we turn \eref{eq:delayed} to the continuous-time case by replacing $n T_\mathrm{s} \to t$\footnote{To turn back to the discrete-time case we can simply sample this expression at $t = n T_\mathrm{s}$ again which will cause aliasing in the spectral estimate. We do not include this effect in the modelling as it is subdominant.}, set $d = d(t)$ and Taylor-expand in the small quantity $\dot d (t - t_0)$. We recognize that we can decompose $\bar y(t)$ into several components
\begin{equation}
    \bar y(t) = \bar y_\mathrm{const}(t) + \bar y_\mathrm{cont}(t) + \bar y_\mathrm{disc}(t) , \label{eq:delayed_decomposed}
\end{equation}
where,
\begin{align}
    y_\mathrm{const}(t) &= \sum_{m=-\infty}^\infty x\!\left(t - m T_\mathrm{s}\right) k\!\left(m T_\mathrm{s} - d_0\right) , \\
    \bar y_\mathrm{cont}(t) &= \sum_{m=-\infty}^\infty \sum_{q=1}^\infty x\!\left(t - m T_\mathrm{s}\right) \bar k^{(q)}\!\left(m T_\mathrm{s} - d_0\right) \frac{\left(- \dot d \cdot (t - t_0)\right)^q}{q!} , \\
    \bar y_\mathrm{disc}(t) &= -\sgn\!\left(t - t_0\right) \sum_{m=-\infty}^\infty \sum_{q=1}^\infty x\!\left(t - m T_\mathrm{s}\right) \Delta k^{(q)}\!\left(m T_\mathrm{s} - d_0\right) \frac{\left(- \dot d \cdot (t - t_0)\right)^q}{q!} .
\end{align}
The first component $y_\mathrm{const}(t)$ represents the zeroth order term in the expansion and is equivalent to a constant delay (cf. \eref{eq:delayed} with $d = d_0$). The remaining constituents $\bar y_\mathrm{cont}(t)$ and $\bar y_\mathrm{disc}(t)$ are small (higher order in $\dot d$) corrections that capture the time-varying character of the delay. Here, the former capture the smooth (or continuous) components in time and the latter the discontinuous ones. Here, we recognize the $\sgn$-function, which is defined as the sign of its argument (the exception is $\sgn(0) = 0$). Furthermore, the functions $\bar k^{(q)}(\tau)$ and $\Delta k^{(q)}(\tau)$ characterize the offset and the magnitude of the discontinuities in the $q$-th derivative of the kernel function. They are formally defined as,
\begin{align}
    \bar k^{(q)} (\tau) &= \frac{k_+^{(q)}(\tau) + k_-^{(q)}(\tau)}{2} ,\\
    \Delta k^{(q)} (\tau) &= \frac{k_+^{(q)}(\tau) - k_-^{(q)}(\tau)}{2} ,
\end{align}
where $k_\pm^{(q)}(\tau) = \lim_{\epsilon \rightarrow 0^+} k^{(q)}(\tau \pm \epsilon)$ are the derivatives approached from the right and left.

%We will focus on the discontinuous components as those are responsible for the glitch-like feature in \gls{psd} estimates.

Let us now study the contribution of each component in \eref{eq:delayed_decomposed} to an estimate of the \gls{psd}; the periodogram. For a general time-series $z(t)$ of length $T$, the periodogram is defined as $T |Z(f)|^2$. Here, $Z(f)$ is the windowed Fourier transform of $z(t)$ (centred on $t_0$), formally given as
\begin{equation}
    Z(f) = \frac{1}{T} \int_{-T/2}^{T/2}\! w(t) \cdot z(t + t_0) \cdot e^{-2 \pi i f t} \,\mathrm{d}t.
\end{equation}
The window function $w(t)$ must be chosen appropriately to limit the naturally occurring spectral leakage. Throughout this study we use a Kaiser window~\cite{Nutall:1981} with $\beta = 30$ that possesses a sidelobe attenuation of \qty{280}{dB}. As a result of the stochastic nature of laser noise, the periodogram $T |Z(f)|^2$ is a random variable itself. Therefore, to make predictions, we work with its expectation value
\begin{align}
    \mathcal{S}_z(f) &\equiv \E\left\{T \left|Z(f)\right|^2\right\} , \label{eq:psd_expectation} \\
    &\sim \frac{1}{T} \left[|\tilde w(f')|^2 * S_z(f')\right](f) , \label{eq:psd_stationary}
\end{align}
denoted with a calligraphic $\mathcal{S}$ to distinguish it from the true power spectral density $S_z(f)$. If $z(t)$ is a stationary random process the second line holds where we make use of the fact that the Fourier transform of the auto-correlation function $R_z(\tau) = \E\{z(t) \cdot z(t + \tau)\}$ is equal to $S_z(f)$. Here, we can observe natural spectral leakage as the true \gls{psd} $S_z(f)$ being viewed through the spectral window $|\tilde w(f')|^2$. This is the case for the constant-delay component $y_\mathrm{const}(t)$ whose (expected) periodogram yields
\begin{equation}
    \mathcal{S}_{\bar y_\mathrm{const}}(f) = \frac{1}{T} \left[|\tilde w(f')|^2 * \left(\left|\tilde h\!\left(f; d_0\right)\right|^2 \cdot S_x(f)\right)\right](f) , \label{eq:psd_const}
\end{equation}
and thus is equivalent to $|\tilde h(f; d_0)|^2 \cdot S_x(f)$ up to natural spectral leakage (similar to \eref{eq:interpolation_error_psd}).

On the contrary, the components $\bar y_\mathrm{cont}(t)$ and $\bar y_\mathrm{disc}(t)$ in \eref{eq:delayed_decomposed} are non-stationary. Therefore, the equality in \eref{eq:psd_stationary} does not hold any more. We reevaluate \eref{eq:psd_expectation} for the discontinuous component $\bar y_\mathrm{disc}(t)$ as the $\sgn$-function gives rise to additional spectral leakage. We find
\begin{equation}
    \mathcal{S}_{\bar y_\mathrm{disc}}(f) \approx \frac{\dot d^{2 \hat q}}{T} \left[ \left|\tilde v_{\hat q}(f')\right|^2 * \left(\left|\Delta\tilde h_{\hat q}(f')\right|^2 \cdot S_x(f')\right) \right](f) , \label{eq:glitch_leakage}
\end{equation}
where $\tilde v_q(f)$ denotes the Fourier transform of the modified window function defined as
\begin{equation}
    v_q(t) = w(t) \cdot \sgn(t) \cdot \frac{t^q}{q!} ,\label{eq:modified_window}
\end{equation}
and $|\Delta\tilde h_q(f)|^2$ is a transfer-function-like factor given as
\begin{equation}
    |\Delta\tilde h_q(f)|^2 = \left|\sum_m \Delta k^{(q)}\!\left(m T_\mathrm{s} - d_0\right) \cdot e^{-2 \pi i f m T_\mathrm{s}}\right|^2 .
\end{equation}
In \eref{eq:glitch_leakage} we limit ourselves to only the leading order contribution in $\dot d$, i.e., the degree $q = \hat q$ of the lowest-order derivative that is discontinuous. For $q < \hat q$ the derivatives $k^{(q)}(\tau)$ are actually continuous and hence $|\Delta\tilde h_q(f)|^2$ vanishes. Contributions from higher degrees $q > \hat q$ are higher order in $\dot d$ and are therefore negligible.

\begin{figure}
    \centering
    \includegraphics{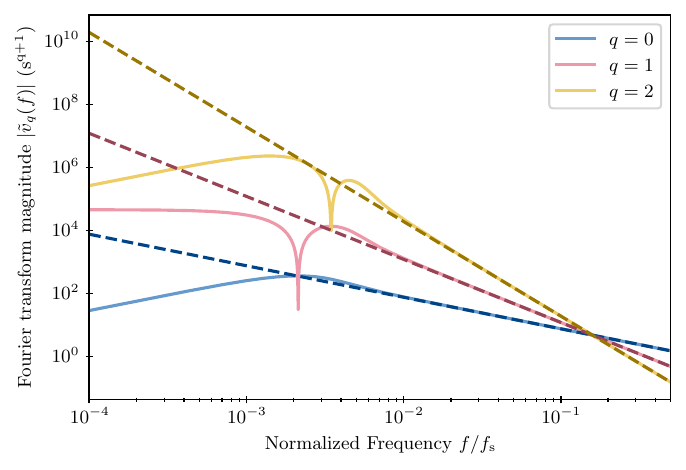}
    \caption{One-sided representation of the modified spectral window magnitude $|\tilde v_q(f)|$ assuming a Kaiser window with $\beta$ of \num{30} and an observation time $T$ of \qty{10000}{\s}. Results for $q=0$ (blue), $q=1$ (red) and $q=2$ (yellow) are shown. The dashed lines show the limit $f \gg \frac{1}{T}$.}
    \label{fig:leakage_appendix}
\end{figure}

The derived model in \eref{eq:glitch_leakage} is similar to \eref{eq:psd_const}, however, the original spectral window $\tilde w(f)$ is exchanged by the modified spectral window $\tilde v_{\hat q}(f)$ and the \gls{psd} of the original time series is adjusted by the transfer function $|\Delta\tilde h_{\hat q}(f)|^2$. We plot the modified window $\tilde v_q(f)$ in \fref{fig:leakage_appendix} for different degrees $q$. We observe weak attenuation at frequencies larger than the frequency resolution $\Delta f = \frac{1}{T}$ where the spectral windows take the form $|\tilde v_q(f)| \to \frac{2 |w(0)|}{(2 \pi f)^{q + 1}}$ and thus only decay as $f^{-(q+1)}$.

%Furthermore, this additional leakage is proportional to increasing powers of the delay derivative $\dot d$ and depends on the transfer function, $|\Delta\tilde h_q(f)|^2$, which is dependent on the interpolation kernel. 

In summary, the magnitude of additional leakage due to an interpolation glitch is strongly dependent on the smoothness of the interpolation kernel. The more continuous derivatives the kernel function has (up to some degree $\hat q$) the weaker the impact of additional leakage and the stronger the suppression due to the factor $\dot d^{2 \hat q}$. Moreover, the magnitude of the discontinuity $\Delta k^{(q)}(\tau)$ plays a role. Through out this study we choose parameters representative of the \gls{lisa} mission. We assume a maximum delay derivative of $\dot d = \num{e-7}$ which is a typical value for the round-trip delay in \gls{lisa} (applied in the $\sigma$ variable, see \eref{eq:sigma}) and a duration of $T = \qty{10000}{\s}$ to enable spectral resolution at the low-frequency end of the \gls{lisa} band, i.e., \qty{0.1}{\milli\Hz}. We will discuss the expected levels of additional leakage for Lagrange interpolation and the cosine-sum kernel in the forthcoming sections.

% also talk about critical delay already
% connect the smoothness and Fourier transform decay argument, do we really need this argument anymore?

\section{Lagrange Interpolation}
\label{sec:lagrange_interpolation}

\begin{figure}
    \centering
    \includegraphics{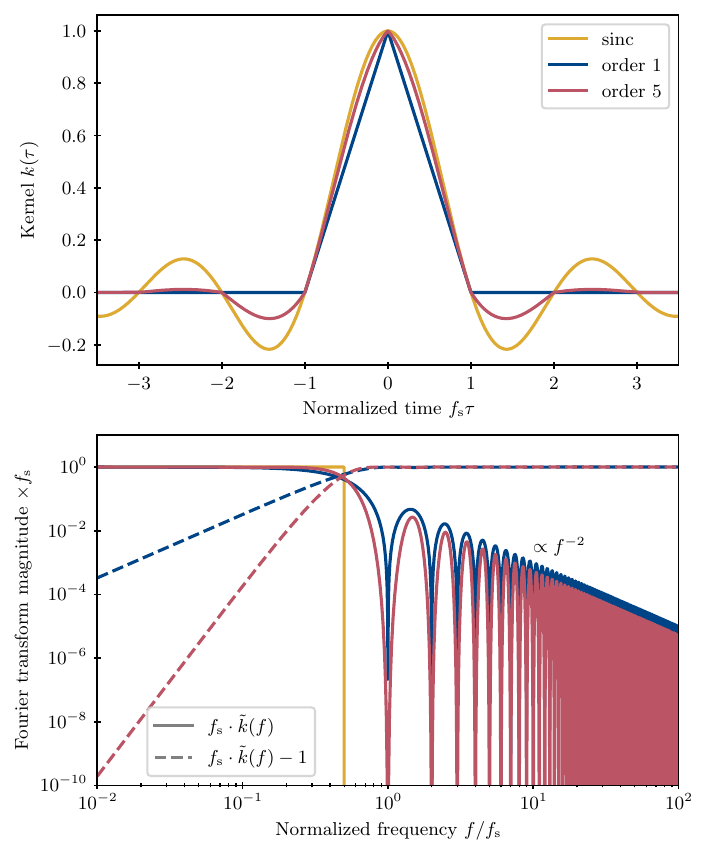}
    \caption{Lagrange kernel functions (upper panel) and their corresponding Fourier transform (lower panel) for orders 1 (blue) and 5 (red). For comparison, we also plot the result for the $\sinc$-kernel which is the limit for Lagrange kernels of increasing order. In the lower panel we additionally show the difference of $f_\mathrm{s} \cdot |\tilde k(f)|$ from unity as dashed lines.}
    \label{fig:lagrange_kernel}
\end{figure}

\begin{figure}
    \centering
    \includegraphics{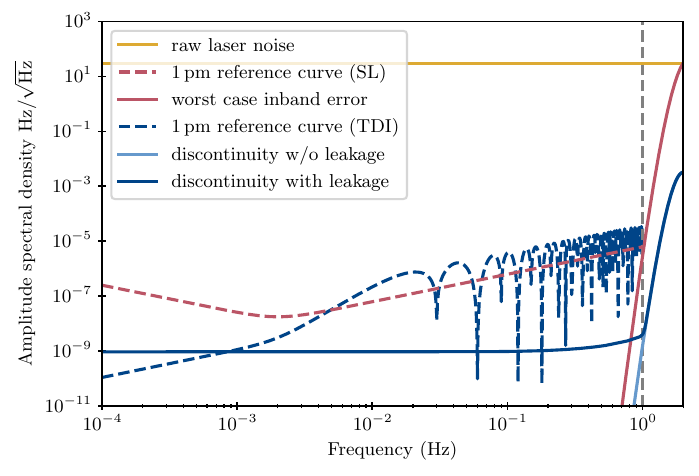}
    \caption{\Glspl{asd} of interpolation errors for Lagrange interpolation order 41 operating on laser noise (in yellow) with an \gls{asd} of \qty{30}{\Hz\per\sqrt{\Hz}}. The light red line shows the worst case interpolation error (see \eref{eq:worst_case_interpolation_error}) which stays below the single link (SL) \qty{1}{\pico\m} reference curve (dashed dark red) in band. An estimate of the \gls{asd} caused by the discontinuity of $k'(\tau)$ (i.e. $q=1$) is shown in blue with (dashed-dotted line) and without leakage (solid line). Here, we assume a duration of $T = \qty{10000}{\s}$, a delay derivative of $\dot d = \qty{e-7}{}$ and a Kaiser window with $\beta = 30$. We note that in-band frequencies are dominated by leakage, which violates the \qty{1}{\pico\m} reference curve after \gls{tdi}~\cite{Staab:2023qrb} (dashed blue).}
    \label{fig:lagrange_glitch}
\end{figure}

% we limit ourselves to N even
The most common interpolation method for \gls{tdi} in \pgls{lisa} is Lagrange interpolation~\cite{Shaddock:2004ua}. It provides excellent low-frequency performance, as its response in the Frequency domain is maximally flat at DC. Lagrange interpolation is based on the Lagrange polynomial which fits $N$ data points using an $(N - 1)$-th degree polynomial. Assuming that data points are equally spaced by the sampling time $T_\mathrm{s}$ and we always choose the data points such that the resulting polynomial is evaluated close to the barycentre of the points, we arrive at the following piece-wise definition of the kernel function $k(\tau)$. For $m = 0, \ldots, N/2 - 1$ we define
\begin{equation}
    k_m(\tau) = \prod_{\substack{n=-N/2\\n\neq m}}^{N/2 - 1} 1 + \frac{|f_\mathrm{s} \tau|}{n - m} \qquad \text{valid for $m \le |f_\mathrm{s} \tau| < m + 1$} . \label{eq:lagrange_kernel}
\end{equation}
%To evaluate $k(\tau)$ we have to choose $m$ such that the condition $m \le |f_\mathrm{s} \tau| < m + 1$ holds.
For $|f_\mathrm{s} \tau| \ge N / 2$ the kernel function is set to zero.

In the upper panel of \fref{fig:lagrange_kernel} we plot the kernel function $k(\tau)$ for polynomial order 1 ($N = 2$) in blue and 5 ($N = 6$) in red. The piece-wise polynomial definition becomes apparent as $k(\tau)$ exhibits non-smooth transitions at integer multiples of the sampling time $T_\mathrm{s}$. For increasing order the kernel function converges to the $\sinc$-function (in yellow), which represent ideal band-limited interpolation.

To assess the in-band performance of Lagrange interpolation we derive the Fourier transform of the kernel $\tilde k(f)$ which is of the form
\begin{equation}
    \tilde k(f) = \frac{1}{f_\mathrm{s}} \cdot P\!\left((\pi f / f_\mathrm{s})^2\right) \cdot \sinc^N\!(f / f_\mathrm{s}) , \label{eq:lagrange_ft}
\end{equation}
where $P(x)$ is a polynomial of degree $N / 2 - 1$. In the lower panel of \fref{fig:lagrange_kernel} we plot $\tilde k(f)$ for orders 1 and 5. Here, we can clearly observe the maximal-flatness property of the Lagrange kernel as at low frequencies the difference of $f_\mathrm{s} \cdot \tilde k(f)$ from unity follows a power law of order $N$ (therefore, derivatives of $\tilde k(f)$ evaluated at $f = \qty{0}{\Hz}$ up to order $N - 1$ are vanishing). At frequencies past the Nyquist frequency, regardless of the interpolation order, the magnitude of $\tilde k(f)$ decays with $f^{-2}$ as a result of the discontinuous first derivative of the kernel function. At first sight, this slow decay seems problematic as it produces large aliased contributions when evaluating the in-band interpolation error (see \eref{eq:worst_case_interpolation_error}). However, since $\tilde k(f)$ quickly drops to zero at integer multiples of $f_\mathrm{s}$ (due to the $\sinc$-function in \eref{eq:lagrange_ft}) which are aliased to dc the impact is greatly reduced and most of the high-frequency content of $\tilde k(f)$ is folded to frequencies close to the Nyquist frequency.

To reduce the impact of interpolation errors below \qty{1}{\Hz} in \pgls{lisa} we require a minimum order of 41~\cite{Staab:2023qrb}. We illustrate this in \fref{fig:lagrange_glitch} where we plot the worst case in-band interpolation error (red) for any constant delay for the usual level of white laser frequency noise with a level of \qty{30}{\Hz\per\sqrt{\Hz}} (yellow). For in-band frequencies, the interpolation error is strictly below the single-link \qty{1}{\pico\m} reference curve (dashed red).

Problems arise when considering a time-varying delay. As discussed in the previous section, discontinuities in derivatives of the kernel function $k^{(q)}(\tau)$ give rise to prominent time-domain features that introduce additional power in \gls{asd} estimates. Taking a closer look at \eref{eq:lagrange_kernel} reveals that odd derivatives of $k(\tau)$ are discontinuous at $f_\mathrm{s} \tau = -\frac{N}{2}, \ldots, \frac{N}{2}$. Therefore, the critical time of a glitch occurrence is when the delay becomes an integer multiple of the sampling time $T_\mathrm{s}$.

We estimate the impact of such an event in \gls{tdi} under the worst-case assumption that the critical condition is met in the final delay operation when calculating the intermediary variable $\sigma_{ij}$ (see \eref{eq:sigma}). In this case the glitch is additive in the final Michelson combination without the usual TDI transfer function applied that would suppress power at low frequencies. We use \eref{eq:glitch_leakage} to estimate the contribution to the \gls{psd} of the discontinuity in the first derivative ($\hat q = 1$) of the kernel function.

In \fref{fig:lagrange_glitch} we plot the result of this computation as the solid blue line. To distinguish spectral leakage and its origin we also consider the case without leakage (light blue) treating the spectral window like a Dirac-delta function\footnote{Note, that we cannot design the window function $w(\tau)$ such that we achieve this behaviour. Thus, this is rather a hypothetical scenario used for illustrative purposes.}. This provides us with a scaled version of $|\Delta \tilde h_q(f)|^2 \cdot S_x(f)$ and, therefore, with a sense for the origin of the leaked power. We note that for \pgls{lisa} in-band frequencies the leaked contribution from out-of-band frequencies, i.e. \qtyrange{1}{2}{\Hz}, dominates. Thus, suppressing laser frequency noise close to the Nyquist rate using a low-pass filter applied prior to \gls{tdi} would greatly reduce the impact of the interpolation glitch for the specific case of Lagrange interpolation. Note that such an additional filter would require a large number of coefficients to realize a sharp transition at \qty{1}{\Hz} (i.e. a close-to-unity response in the \gls{lisa} band and appropriate suppression between \qty{1}{\Hz} and \qty{2}{\Hz}). This comes with a few drawbacks, e.g., the size of data gaps is increased and the introduction of an additional group delay.

\section{Cosine-sum Kernel}
\label{sec:cosine-sum_kernel}

\begin{figure}
    \centering
    \includegraphics{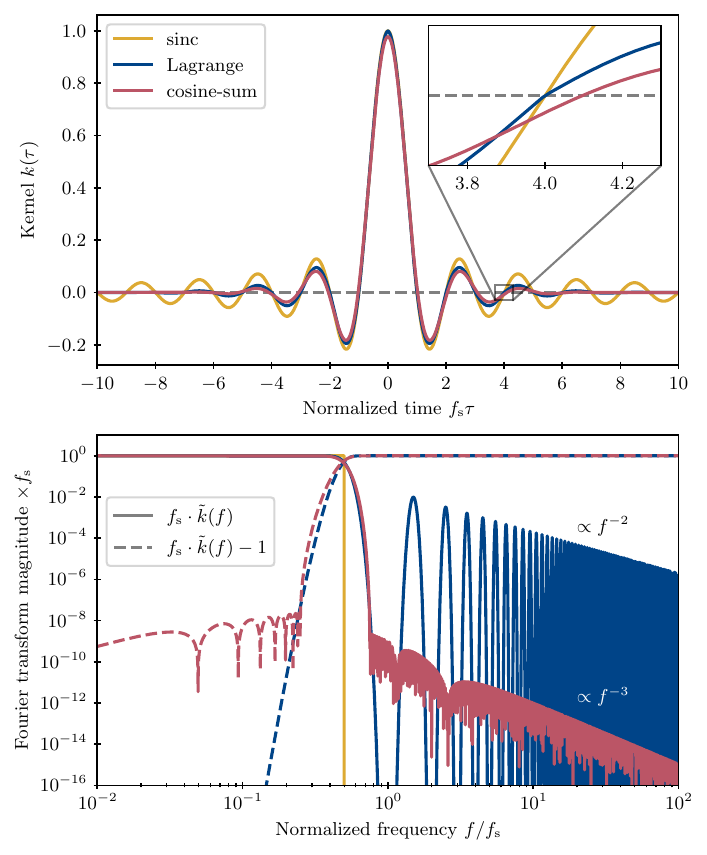}
    \caption{Comparison of the ideal $\sinc$-kernel in yellow, the order 41 Lagrange kernel ($N=42$) in blue and the cosine-sum kernel ($N=22$) in red. The upper panel presents the time-domain kernel functions and their different behaviours at integer times (highlighted in the zoomed-in insert). Both the $\sinc$-kernel and Lagrange kernel are not displayed completely as they reach past the x-axis' limits. The lower panel shows the normalized Fourier transform in solid and its difference from unity in dashed.}
    \label{fig:cosine-sum_kernel}
\end{figure}

\begin{figure}
    \centering
    \includegraphics{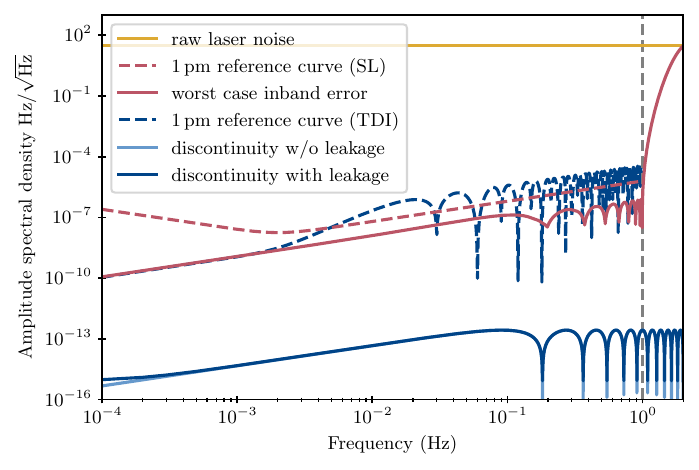}
    \caption{\Glspl{asd} of interpolation errors for the cosine-sum kernel with otherwise identical parameters as \fref{fig:lagrange_glitch}. For this choice of kernel both the worst-case in-band interpolation error (red) and the \gls{asd} estimate of the leading order discontinuity (blue), i.e. in the second derivative ($\hat q=2$), stay below their respective \qty{1}{\pico\m} reference curve (in dashed). The modeled \gls{asd} estimates of the discontinuity with and without leakage are almost identical, indicating that additional in-band power is not leakage-dominated.}
    \label{fig:cosine-sum_glitch}
\end{figure}

% Lagrange interpolation is not ideal for LISA
% 1) relatively high number of coefficients
% 2) promiment interpolation glitch for realistic LISA setup
% solution: design interpolation kernels that optimize for LISA in-band frequencies and thus use less coefficients + kernel that is more flat

The short-comings of Lagrange interpolation call for an alternative interpolation method. In this section, we propose a novel interpolation kernel which allows us to reduce the width of $k(\tau)$ (and therefore a reduction of interpolation coefficients) and, simultaneously, suppress the impact of discontinuities in the interpolation kernel in \gls{tdi}. To tackle these aims, we explore kernel functions that are derived from a finite Fourier series that only consider the even cosine contributions. For this reason, we dub this novel family of kernels ``cosine-sum kernels''. They are defined as
\begin{equation}
    k(t) = \rect\!\left(\frac{t}{N T_\mathrm{s}}\right) \sum_{n=0}^{N-1} a_n \cdot \cos\!\left(2 \pi f_\mathrm{s} \frac{n}{N} t\right) , \label{eq:cosine_sum}
\end{equation}
where $N T_\mathrm{s}$ is the width of the kernel (and therefore $N$ the number of coefficients drawn from it) and $a_n$ the Fourier coefficients.

This formulation of the kernel function is highly versatile and in order to fulfil the constraints of the kernel design the coefficients $a_n$ have to be determined. For an effective kernel design for the constant delay operation we require two conditions which we derive from \eref{eq:worst_case_interpolation_error}. Firstly, the in-band response (\qtyrange{0}{1}{\Hz}) of the (normalized) Fourier transform $f_\mathrm{s} \cdot \tilde k(f)$ should be close to unity and, secondly, the out-of-band response ($f \ge \qty{3}{\Hz}$) should be well attenuated to suppress the aliased contributions in \eref{eq:worst_case_interpolation_error}. Those conditions are cast into a minimax optimization problem that can be solved iteratively using a variant of the Parks-McClellan algorithm~\cite{Parks:1972}. By defining appropriate weights for the error function we are able to design a kernel function that accounts for the $f$-slope of the \qty{1}{\pico\m} reference curve (in frequency units). The result is presented in \ref{app:optimization} and the accompanying figure therein.

As derived in \sref{sec:interpolation} we require a sufficiently smooth kernel function to avoid the introduction of glitch-like features in the data. The family of cosine-sum kernels in \eref{eq:cosine_sum} is infinitely differentiable almost everywhere except at the boundaries $\tau = \pm N T_\mathrm{s} / 2$ where even derivatives are discontinuous. To control the effect of those discontinuities (see \eref{eq:glitch_leakage}) we also include the possibility to increase the smoothness of $k(\tau)$ by imposing additional conditions on the $a_n$ such that $k^{(2q)}(\tau) = 0$ with $q \ge 0$ at the boundaries. This comes at the cost of degrading the ideal response of $\tilde k(f)$ as some degrees of freedom have to be invested in fulfilling those boundary conditions. This is discussed in more detail in \ref{app:optimization}.

Our final proposal uses $N = 22$ coefficients and assures continuity in the first derivative (one additional degree of smoothness compared to the Lagrange kernel). The resulting kernel function $k(\tau)$ and its Fourier transform are presented in \fref{fig:cosine-sum_kernel}. The kernel functions for Lagrange interpolation (blue) and of the cosine-sum family (red) are almost identical. However, on closer inspection there are some notable differences highlighted by the zoomed-in insert. First, opposed to the ideal $\sinc$-function and the Lagrange kernel the cosine-sum kernel is non-unity at $\tau = 0$ and is lacking zero crossings at $f_\mathrm{s} \tau \in \mathbb{Z}$ (with relative differences of up to a few percent). As a result, the cosine-kernel does not strictly constitute an interpolator as the interpolating function in \eref{eq:interpolation} does not cross the original samples\footnote{In other words, applying the delay operation with $d=0$ does not yield the original time series.}. Furthermore, the cosine-sum kernel appears smooth over the entire domain (except the boundaries) as discussed already before.

Finally, let us investigate the in-band performance of the cosine-sum kernel for constant and time-varying delays. Analogously to the previous section, in \fref{fig:cosine-sum_glitch}, we present the worst-case interpolation error (red) and the \gls{asd} estimate of the discontinuity (blue) in the second derivative ($\hat q = 2$). Both stay below their \qty{1}{\pico\m} reference curve (single-link and \gls{tdi}-adjusted~\cite{Staab:2023qrb}, respectively). We also note that the \gls{asd} estimates for the discontinuity are not leakage-dominated as $\Delta \tilde h_{\hat q}(f)$ is not peaked at high frequencies as it is the case for the Lagrange kernel, so prior filtering of the raw measurements is not required.

\section{Numerical Simulations}
\label{sec:simulation}

\begin{figure}
    \centering
    \includegraphics{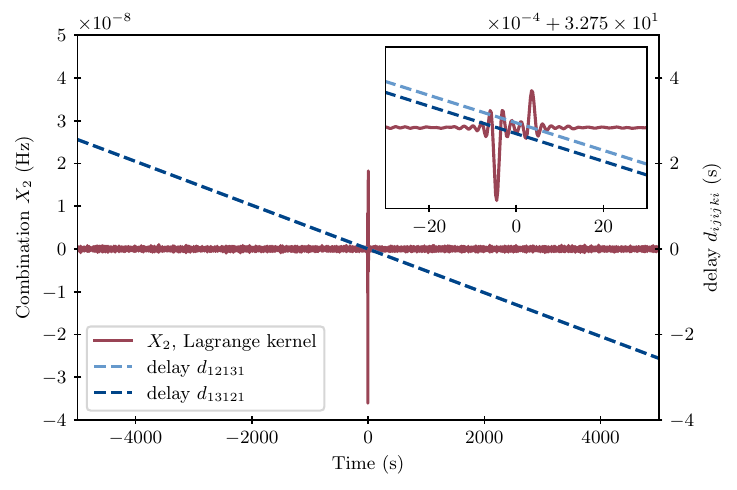}
    \caption{Second-generation Michelson combination $X_2$ computed using Lagrange interpolation (order \num{41}). In dark red the low-pass filtered time series is shown exhibiting a prominent glitch at the centre. The dashed lines represent the round-trip delays $d_{12131}$ (light blue) and $d_{13121}$ (dark blue) which are almost identical. The inset shows a zoomed-in view which reveals that the time series of $X_2$ contains two distinct features that happen when each of the round-trip delays crosses the numerical value \qty{32.75}{\s} (an integer multiple of the sampling time $T_\mathrm{s} = \qty{0.25}{\s}$).}
    \label{fig:lisa_glitch}
\end{figure}

\begin{figure}
    \centering
    \includegraphics{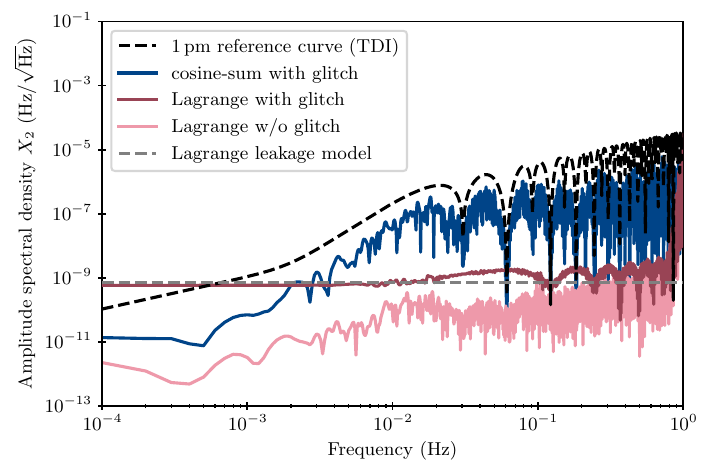}
    \caption{\Gls{asd} of second-generation Michelson combination $X_2$ for different interpolation schemes. Periodograms of \qty{10000}{\s}-datasets including the critical time (c.f. \fref{fig:lisa_glitch}) are shown for numerical \gls{tdi} calculations using the cosine-sum kernel (dark blue) and Lagrange interpolation (dark red). In light red we plot the result for Lagrange interpolation for a time slice of the data that excludes the glitch. For reference, we plot the level of additional leakage at DC due to an interpolation glitch in dashed grey and the \qty{1}{\pico\m} reference curve (propagated through \gls{tdi}~\cite{Staab:2023qrb}) in dashed black.}
    \label{fig:lisa_performance}
\end{figure}

In the following, we demonstrate the impact of the choice of interpolation kernel in \gls{tdi} for a realistic \pgls{lisa} setup. Therefore, we run numerical simulations using LISA Instrument~\cite{bayle_2024_13809621}. We simulate interferometric data at \qty{4}{\Hz} for \qty{10000}{\s} to resolve down to the band limit of \qty{e-4}{\Hz} only including white laser frequency noise at a level of \qty{30}{\Hz\per\sqrt{\Hz}}. Furthermore, to simplify the setup we assume a single laser per spacecraft to directly yield the $\eta$ variable (see \eref{eq:eta}) as an output of the simulation. For realistic time-varying delays we choose numerical orbit files provided by ESA~\cite{Martens:2021phh} and pick the starting time of the simulation such that the round-trip delays $d_{12131}$ and $d_{13121}$ become exactly integer delays in the centre of the domain. When calculating the second-generation Michelson variable $\dot X_2$\footnote{As LISA Instrument simulates interferometric data in units of frequency the expressions in \eref{eq:x2} need to be adjusted by replacing each delay operator by its Doppler shift corrected equivalent~\cite{Bayle:2021mue}. The result of this calculation is the Michelson combination in units of frequency indicated by the dot.} as defined in \eref{eq:x2} using PyTDI~\cite{staab_2023_8429119} the interpolation glitch for the kernel functions discussed in the previous section enters through the calculation of the intermediary variable $\sigma$. This presents the worst case coupling, as the effect of the glitch is not attenuated by the transfer function of \gls{tdi} that usually acts like a time derivative, thus suppressing low-frequency components.

In \fref{fig:lisa_glitch} we show the time-domain representation of the Michelson variable $X_2$ (in red) using Lagrange interpolation. After applying a low-pass filter to suppress the dominant contributions to the interpolation error at high frequencies the interpolation glitch becomes clearly visible in the centre of the domain when the round-trip delays $d_{12131}$ (dashed light blue) and $d_{13121}$ (dashed dark blue) are equal to \qty{32.75}{\s} which is an integer multiple of the sampling time of \qty{0.25}{\s}. When zooming in on the glitch it separates into two distinct features which can be unambiguously attributed to the integer-delay crossing of each of the round-trip delays.

To circumvent this strong interpolation glitch we compute the same \gls{tdi} variable using the cosine-sum kernel introduced in \sref{sec:cosine-sum_kernel}. As it disappears in the time-domain result we resort to the spectral representation to check for anomalies. In \fref{fig:lisa_performance} we plot the periodograms (using the same Kaiser window as before with $\beta = \num{30}$) of the Michelson variables computed using the cosine-sum kernel (blue) and Lagrange interpolation (dark red). As expected, the cosine-sum kernel performs closer to the \qty{1}{\pico\m} reference curve~\cite{Staab:2023qrb} than Lagrange interpolation for most of the band. However, the latter contains extra power that is not explained by the usual (constant delay) in-band interpolation error. We confirm this by repeating the experiment using Lagrange interpolation for data shifted to a slightly later time to exclude the glitch (light red). The level of the extra power in the experiment including the glitch (dark red) is consistent with the model (grey) derived in \sref{sec:lagrange_interpolation} assuming $\dot d$ has a magnitude of approximately \num{5e-8}. We calculate the level of spectral leakage at DC due to an interpolation glitch using \eref{eq:glitch_leakage} assuming $S_x(f) \to S_{\dot \rho}(f) = 4 \sin^2(4\pi f L) \cdot \big(\qty{30}{\Hz\per\sqrt{\Hz}}\big)^2$ with $L = \qty{8.19}{\s}$. The oscillatory features visible at higher frequencies can be explained by the fact that the pair of interpolation glitches enter with a separation of approximately \qty{10}{\s} which explains the nulls at multiples of \qty{0.1}{\Hz}.

\section{Conclusion}
\label{sec:conclusion}

Interpolation is an integral function of \gls{tdi} which is needed to suppress laser frequency noise in \gls{lisa}. \Gls{tdi} forms equal-arm interferometers by building linear combinations of time-shifted beatnote phases of heterodyne interferometers. The time-shift operation requires an extremely accurate interpolation method to cancel out laser phases and suppress laser frequency noise by up to nine orders of magnitude. In the past, Lagrange interpolation was used~\cite{Shaddock:2004ua} for this task. However, we found that it produces glitch-like features occurring when time-varying delays become integer multiples of the sampling time.

In this paper we develop a generalized framework to describe interpolation errors arising when time-shifting discrete-time data samples. We limit ourselves to interpolation methods that are defined as a convolution of a kernel function with the data to yield a continuous representation. For any given constant delay, the in-band interpolation error can be directly calculated from the kernel function. In case of a time-varying delay, the delay operation becomes time-variant and, thus, cannot be represented as a linear time-invariant system any more. The consequence is spectral leakage, which is contaminating the \gls{lisa} band with excess power. The level of spectral leakage is dependent on the rate of change of the delay and the degree of smoothness of the kernel function.

Lagrange interpolation has been the standard interpolation method for \gls{tdi} since its inception. Previous investigations of interpolation method performance only considered constant time delays~\cite{Shaddock:2004ua}. In this paper, we demonstrate that Lagrange interpolation for a realistic LISA configuration is insufficient as the effect of spectral leakage rises above the \qty{1}{\pico\m} reference curve at low frequencies. Here, we assume a delay derivative of the order of \num{e-7}, which is a typical value for the round-trip delay in the Michelson combination ($d_{12131}$ and $d_{13121}$), and choose an observation time of \qty{10000}{\s} to resolve down to the low-end of the \gls{lisa} band. The reason for the high level of spectral leakage is the discontinuity of the Lagrange kernel function in its first derivative.

As a solution, we propose a new family of versatile kernel functions composed of a sum of cosine functions (see \eref{eq:cosine_sum}). We describe a method to efficiently optimize the free parameters of the cosine-sum kernel to yield a sufficient in-band performance and a high degree of smoothness. At the same time, we aim to minimize the width of the kernel (which translates directly into the number of interpolation coefficients) to reduce the cost of the delay operation and be more resilient to boundary effects. We find that a cosine-sum kernel that possesses a continuous first derivative and yields 22 interpolation coefficients suffices to achieve adequate in-band interpolation errors and suppress additional spectral leakage. This constitutes a reduction in number of interpolation coefficients by roughly a factor of two. For reference, we print the 22 coefficients that define the optimized cosine-sum kernel in \tref{tab:coefficients}.

To numerically check the performance of both interpolation kernels we run \gls{lisa} simulations assuming realistic dynamics based on orbit files provided by ESA. We intentionally choose a time frame of the mission where the round-trip delays $d_{12131}$ and $d_{13121}$ become integer multiples of the sampling time of the data. This results in a prominent glitch in the time-domain when calculating the second-generation Michelson variables in their factorized form, which in turn leads to strong spectral leakage. We then evaluate the periodogram for the \gls{tdi} variable $X_2$ obtained using both kernels, the Lagrange and the optimized cosine-sum kernel. The \gls{asd} estimate of the former is affected by spectral leakage that is consistent with the levels predicted by the derived model. On the contrary, the periodogram of $X_2$ evaluated using the cosine-sum kernel is free of excess power.

We therefore suggest using the cosine-sum kernel instead of Lagrange interpolation for the time-shift operation in \gls{tdi}. This choice drastically cuts down the number of interpolation coefficients, and thus the computational cost, by a factor of two and reduces the amount of invalid samples at the edges of the time series or around gaps where the interpolation kernel does not overlap completely with the data. Moreover, the cosine-sum kernel suppresses additional spectral leakage induced by discontinuities in the kernel function to adequate levels.

% make note on using interpolation also for synchronization purposes, how accurate does it need to be for case1/case2/case3?
% locking configuration three lasers (shouldn't have big impact on level, but different locking configs produce different correlations among the lasers and TDI combinations become asymmetric)
% transfer function of the interpolation glitch depends on the factorization of the TDI combination
% only even number of interpolation coefficients

\ack
The authors thank the LISA Simulation Working Group and the LISA Simulation Expert Group for the lively discussions on all simulation-related activities. M.S., A.H., M.L.\ and P.W.\ gratefully acknowledge support by the Centre National d’\'Etudes Spatiales (CNES). M.S. is supported by the research program of the Netherlands Organisation for Scientific Research (NWO).
J-B.B.\ and G.W.\ acknowledge the support of the UK Space Agency, UKSA, through grant ST/Y004906/1.
O.H.\ acknowledges the support of the German Space Agency, DLR.  The work is supported by the Federal Ministry for Economic Affairs and Climate Action based on a decision by the German Bundestag (FKZ 50OQ1801 and FKZ 50OQ2301).

\FloatBarrier
\appendix

\section{Optimizing the Cosine-sum Kernel}
\label{app:optimization}

Using the class of cosine-sum kernels introduced in \sref{sec:cosine-sum_kernel} we can optimize the frequency response of the interpolation kernel. This is achieved by choosing a set of coefficients $\{a_n\}_{n=0}^{N-1}$ that fulfil some optimality criteria. For the interpolation method in \gls{tdi}, we identify two such criteria in \sref{sec:interpolation}; the in-band performance of interpolation and the smoothness of the kernel. In \eref{eq:worst_case_interpolation_error} we relate the Fourier transform of the kernel to the worst-case interpolation error over all possible values of the delay $d$. This formula tells us that $\tilde k(f)$ should be unity inside the band and vanishing for frequencies $|f| > \frac{f_\mathrm{s}}{2}$. 

Therefore, in the first step, we shall derive the Fourier transform of the cosine-sum kernel defined in \eref{eq:cosine_sum}. A similar derivation is given in~\cite{albrecht_tailoring_2010}. We start by expressing the Fourier transform as
\begin{equation}
    \tilde k(f) = \frac{N T_\mathrm{s}}{2} \sum_{n=0}^{N-1} a_n \left(\sinc\!\left(N T_\mathrm{s} f + n\right) + \sinc\!\left(N T_\mathrm{s} f - n\right)\right) , \label{eq:cosine_sum_ft}
\end{equation}
since the cosine transforms to a pair of Dirac-delta functions which shift the argument of the $\sinc$-function (which is the Fourier transform of the rectangular function). Next, we introduce the normalized frequency $\xi = N T_\mathrm{s} f$ for the sake of brevity and rewrite the $\sinc$-function as $\sinc(\xi \pm n) = \frac{(-1)^n \sin(\pi \xi)}{\xi \pm n}$. We arrive at the simplified expression
\begin{align}
    \tilde k(\xi) &= N T_\mathrm{s} \sinc(\xi) \sum_{n=0}^{N-1} a_n (-1)^n \frac{\xi^2}{\xi^2 - n^2} , \\
    &= \underbrace{\frac{\sinc(\xi)}{\prod_{n=1}^{N-1} \xi^2 - n^2}}_{\equiv R(\xi)} \cdot P\!\left(\xi^2\right) , \label{eq:cosine_sum_ft_xi}
\end{align}
which can be recast into a product of the function $R(\xi)$ defined above and a polynomial $P(x)$ in $\xi^2$ of degree $N-1$. This is achieved by contracting the sum of fractions in the first line into a single fraction with a common denominator. The polynomial also absorbs the free parameters $a_n$. Therefore, the optimization problem can be restated as finding the polynomial $P(x)$ such that aforementioned conditions are fulfilled. To relate back to the parameters $a_n$ required to compute \eref{eq:cosine_sum} we notice the following relationship. Let us evaluate the Fourier transform of the kernel at $\xi = m = 0, 1, \dots, N-1$. Comparing the results from equations \ref{eq:cosine_sum_ft} and \ref{eq:cosine_sum_ft_xi} we find
\begin{equation}
    a_m = \frac{(-1)^m}{N T_\mathrm{s}} \cdot \frac{P\!\left(m^2\right)}{\prod_{\substack{n=0 \\ n \neq m}}^{N-1} m^2 - n^2} ,
\end{equation}
making use of l'H\^opital's rule for calculating the limit of $R(\xi)$ for $\xi \rightarrow m$ in \eref{eq:cosine_sum_ft_xi}.

Let us now turn back to the original optimization problem; finding the set of coefficients $\{a_n\}_{n=0}^{N-1}$ defining the cosine-sum kernel that fulfil the aforementioned optimality criteria. We define the weighted error function as
\begin{equation}
    E(\xi) = W(\xi) \left(\tilde k(\xi) - D(\xi)\right) , \label{eq:error_function}
\end{equation}
where $W(\xi)$ denotes frequency dependent weights and $D(\xi)$ the desired response which is unity for frequencies until the pass-band frequency $f_\mathrm{pass}$ and zero from the stop-band frequency $f_\mathrm{stop}$. To obtain the coefficients that respect the inband performance we have to minimize the maximum absolute weighted error over all normalized frequencies $\xi \in U = [0, N T_\mathrm{s} f_\mathrm{pass}] \cup [N T_\mathrm{s} f_\mathrm{stop}, \infty)$
\begin{equation}
    \{a_n\}_{n=0}^{N-1} = \argmin \max_{\xi \in U} |E(\xi)| . \label{eq:minimax}
\end{equation}
The solution to this problem is in general difficult to find. However, in our case we can recast the expression in \eref{eq:error_function}  into
\begin{equation}
    E(\xi) = \underbrace{\vphantom{\frac{D(\xi)}{R(\xi)}}W(\xi) \cdot R(\xi)}_{\hat W(\xi)} \bigg(P\!\left(\xi^2\right) - \underbrace{\frac{D(\xi)}{R(\xi)}}_{\hat D(\xi)}\!\bigg) ,
\end{equation}
by plugging in \eref{eq:cosine_sum_ft_xi} which turns \eref{eq:minimax} into a weighted Chebyshev approximation problem as in~\cite{Parks:1972}. We identify the corresponding weighting function $\hat W(\xi)$ and desired response $\hat D(\xi)$. Furthermore, we require that the approximation is exact at DC, i.e., $E(0) = 0$. Additionally, in the numerical implementation of the algorithm, we simplify the weighting function as
\begin{equation}
    \hat W(\xi) \sim
    \begin{cases*}
        W(\xi) \cdot R(\xi) & if $\xi \le N - \frac{1}{2}$, \\
        \frac{(-1)^{N-1}}{\pi\xi} \frac{1}{\prod_{n=1}^{N-1} \xi^2 - n^2} & else,
    \end{cases*} \label{eq:weighting_approximation}
\end{equation}
to omit the oscillations (and zero-crossings) from the $\sinc$-function that cause numerical difficulties. The approximation is exact for $\xi \le N - \frac{1}{2}$ and else the absolute value presents an upper bound for $|\hat W(\xi)|$. Therefore, it still respects at least the weighting implied by $\hat W(\xi)$. Furthermore, we use an equivalent expression for $R(\xi)$ in \eref{eq:weighting_approximation} to avoid numerical problems close to integer values given as
\begin{equation}
    R(\xi) = \frac{1}{2} \frac{\sum_{n=0}^{N-1} \sinc(\xi + n) + \sinc(\xi - n)}{\sum_{n=0}^{N-1} (-1)^n \left(\prod_{\substack{m=0 \\ m \neq n}}^{N-1} \xi^2 - m^2\right)} .
\end{equation}

Reducing the polynomial degree $M$ of $P(x)$ improves the decay of $\tilde k(f)$ for $f \rightarrow \infty$ (which is equivalent to $\xi \rightarrow \infty$). Let us set $M = N - L$ where $L$ is a positive integer between 1 and $N$. As we apply the limit to \eref{eq:cosine_sum_ft_xi} we find that the Fourier transform decays as $f^{-(2L - 1)}$. The case $L = 1$ yields the slowest decay but allows for many degrees of freedom to shape $\tilde k(f)$. On the contrary, $L = N$ results in $P(x)$ being a constant (only a single polynomial coefficient to tune) but a fast decay proportional to $f^{-(2N - 1)}$. The decay of the Fourier transform is directly linked to the smoothness of the function. A function possesses continuous derivatives up to degree $q$ if its Fourier transform decays at least as $f^{-(q + 2)}$.

\begin{figure}
    \centering
    \includegraphics{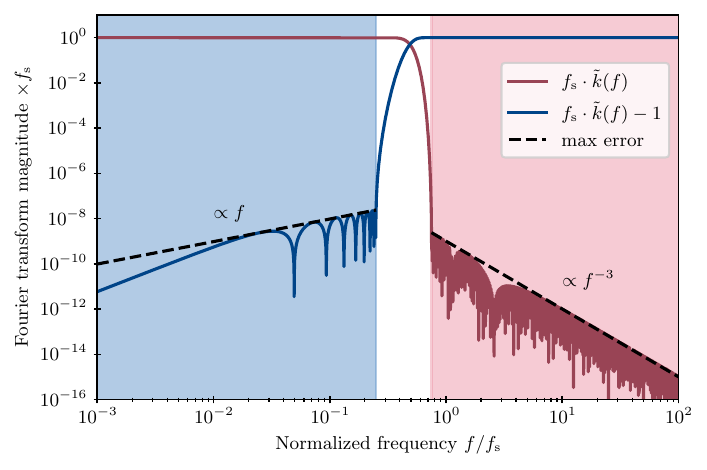}
    \caption{Result of optimization procedure for $N=22$ and $L=2$. The blue shaded region indicates the pass-band (from $0$ to $0.75$) and the red shaded region (from $0.75$ to $\infty$) the stop-band. In dark red the Fourier transform of the kernel is plotted and in dark blue its difference from unity, which represent the (unweighted) error function in the stop-band and pass-band, respectively. In dashed black, we indicate the maximum error resulting from the specific choice of weighting function $W(f)$.}
    \label{fig:kernel_design}
\end{figure}

\begin{table}
    \centering
    \begin{tabular}{c S[table-format = 3.16e2]}
        \toprule
        {Coefficient} & {Value} \\
        \midrule
        $a_{0}$ & 4.5454545454545456e-02 \\
        $a_{1}$ & 9.0909090805559478e-02 \\
        $a_{2}$ & 9.0909091053369862e-02 \\
        $a_{3}$ & 9.0909091301689185e-02 \\
        $a_{4}$ & 9.0909089335187473e-02 \\
        $a_{5}$ & 9.0909089486150965e-02 \\
        $a_{6}$ & 9.0908063258257371e-02 \\
        $a_{7}$ & 9.0809776923836752e-02 \\
        $a_{8}$ & 8.9474437673758789e-02 \\
        $a_{9}$ & 8.2609330488025795e-02 \\
        $a_{10}$ & 6.4821606246711530e-02 \\
        $a_{11}$ & 3.8667853665977497e-02 \\
        $a_{12}$ & 1.5821652446457120e-02 \\
        $a_{13}$ & 3.9863791298782002e-03 \\
        $a_{14}$ & 5.2881589106309470e-04 \\
        $a_{15}$ & 2.6871219117667249e-05 \\
        $a_{16}$ & 1.8192909362438404e-07 \\
        $a_{17}$ & 4.2775612572358636e-11 \\
        $a_{18}$ & 6.4250483670003823e-11 \\
        $a_{19}$ & 1.2734804870086914e-10 \\
        $a_{20}$ & -1.0953461600750713e-10 \\
        $a_{21}$ & 5.2799552835044587e-11 \\
        \bottomrule
    \end{tabular}
    \caption{Optimized coefficients $a_n$ of the cosine-sum kernel defined in \eref{eq:cosine_sum}.}
    \label{tab:coefficients}
\end{table}

In \fref{fig:kernel_design} we illustrate the result of the optimization procedure using parameters that respect the \pgls{lisa} requirements. We assume a sampling frequency of $f_\mathrm{s} = \qty{4}{\Hz}$ and an upper band edge of \qty{1}{\Hz} which determines $f_\mathrm{pass} / f_\mathrm{s} = 1 / 4$ and $f_\mathrm{stop} / f_\mathrm{s} = 3 / 4$. The stop-band frequency follows from arguments of aliasing as the band $[0.5, 0.75]$ is aliased to $[0.25, 0.5]$ which is out-of-band. For sufficient in-band and out-of-band suppression we choose a minimum kernel width of $N=22$. To obtain continuity in the first derivative of the kernel (decay of $f^{-3}$) we set $L=2$. The weighting function $W(f)$ is defined as
\begin{equation}
    W(f) = 
    \begin{cases*}
        (f + f_\mathrm{min})^{-1} & if $0 \le f \le f_\mathrm{pass}$, \\
        10 \cdot f_\mathrm{pass}^{-1} \cdot (f / f_\mathrm{stop})^3 & else,
    \end{cases*} \label{eq:weightings}
\end{equation}
where $f_\mathrm{min}$ is set to \num{0.25e-4} and saturates the weighting function at low frequencies.

Consequently, the maximum error $\delta / W(f)$ (where $\delta = \max_{\xi \in U} |E(\xi)|$) shown in dashed black in \fref{fig:kernel_design} is proportional to the Fourier frequency in band which is desired to respect the \pgls{lisa} requirement curve, which exhibits the same frequency dependency. The weighting function of the interval $[f_\mathrm{stop}, f_\mathrm{s}]$ is set such that $\tilde k(f)$ follows the general decay of $f^{-3}$ and is at least ten times smaller at $f_\mathrm{pass}$ than its difference from unity at $f_\mathrm{pass}$. The latter assures that after "aliasing" (see \eref{eq:worst_case_interpolation_error}) the performance is mostly limited by the in-band error of the kernel function. For reference the \num{22} coefficients are printed in \tref{tab:coefficients}.

\FloatBarrier

\section*{References}
\bibliographystyle{iopart-num}
\bibliography{references}

\end{document}